\documentclass [pra,twocolumn,preprintnumbers,amsmath,amssymb]{revtex4}

\usepackage{graphicx}
\usepackage{dcolumn}
\usepackage{bm}

\DeclareGraphicsExtensions{.eps,.gif,.ps}

\begin{document}

\title{Multi-Bloch Vector Representation of the Qutrit}

\author{Pawe{\l} Kurzy\'nski}
\email{kurzpaw@hoth.amu.edu.pl}

\affiliation{Faculty of Physics, Adam Mickiewicz University,
Umultowska 85, 61-614 Pozna\'{n}, Poland.}


\begin{abstract}
An ability to describe quantum states directly by average values of measurement outcomes is provided by the Bloch vector. For an informationally complete set of measurements one can construct unique Bloch vector for any quantum state. However, not every Bloch vector corresponds to a quantum state. It seems that only for two dimensional quantum systems it is easy to distinguish proper Bloch vectors from improper ones, i.e. the ones corresponding to quantum states from the other ones. I propose an alternative approach to the problem in which more than one vector is used. In particular, I show that a state of the qutrit can be described by the three qubit-like Bloch vectors.
\end{abstract}

\maketitle

\section{Introduction}

A state of a discrete quantum system is described by a complex vector. If only partial information about the system is given, or if it is entangled with any other system, its state is represented by a density matrix. This well known description is always nonintuitive to anyone hearing for the first time of quantum mechanics because it takes place in the abstract Hilbert space instead of the spacetime, in which all physical phenomena occurs. However, some alternative representations of quantum states referring to classical intuition have been introduced. 

An important alternative description of a discrete quantum state is given by the Bloch vector \cite{NC}, which explicitly refers to experimental notions like expectation values of measurements which in principle can be performed in a laboratory. The state is represented by a real vector whose coordinates are simply expectation values of different measurements and whose dimension is given by a minimal number of measurements needed to obtain complete information about the system. Although widely used, the Bloch vector description is still under development \cite{BK,Mend,KZ,KK,Kim,JS}, because, although the idea seems simple, it is hard to provide one important feature. Every quantum state can be written as the Bloch vector, but not every vector corresponds to a quantum state. For some vectors the corresponding {\it density matrix}, though still of trace one, can have negative eigenvalues. 

In Hilbert space of dimension two the set of all Bloch vectors is confined in a real three dimensional ball. Such a nice symmetry has not been found for higher dimensions \cite{BK,Mend,KZ,KK,Kim,JS}. The symmetry allows for an easy recognition of the proper Bloch vectors, i.e., the vectors corresponding to quantum states. However, there is also one more reason to look for well shaped symmetric boundaries. The beauty of the Bloch ball is manifested by the fact that its symmetry not only represents nicely states of the qubit, but also gives very intuitive picture of the dynamics of the system --- every unitary operation is simply a rotation of the ball. Therefore, an additional goal of a generalization of the Bloch vector to higher dimensions is to give an intuitive representation of the dynamics of the underlying system. 

The Bloch vector of the qubit is well know to anyone working in the field of quantum information, optics or magnetism. On the other hand, it is obvious that every quantum system contains a two dimensional (qubit) subspace. In particular, the number of a distinct qubit subspaces of a d-level system is given by a binomial coefficient $\begin{pmatrix} d \\ 2 \end{pmatrix}$. Each of these subspaces can be described by its own qubit Bloch vector. However, these Bloch vectors cannot be independent.

An important extension of the qubit is the qutrit, the three level system. Despite apparent simplicity, the qutrit is complex enough to reveal bizarre properties of quantum theory, like contextuality \cite{KS}. Moreover, it was shown that using qutrits, instead of qubits, one can perform much better in quantum information processing tasks, like for example the quantum key distribution \cite{PP}. 

The aim of this work is to find a set of constraints on Bloch vectors of qubit subspaces of the underlying qutrit. As a result, one obtains a new Bloch-like representation of the qutrit. Moreover, it is going to be shown that this representation is natural for spin 1 whose dynamics is intuitively fitted for this picture. The extension to higher dimensions is also discussed.

\section{Operator Basis}

The usual construction of the Bloch vector for an arbitrary qudit starts with a choice of an appropriate measurements which mathematically are represented by $d^2$ elements of an orthogonal hermitian operator basis. The orthogonality is according to the trace product
$(A,B)=\text{Tr}\{A^{\dagger}B\}$.
Due to the normalization constraint, there are $d^2-1$ parameters characterizing the state of the qudit. Usually \cite{BK,Mend,KZ,KK,Kim,JS}, all but one basis operator, which is just the identity divided by $d$, are chosen to be traceless so any density matrix can be written as
\begin{equation}\label{e1}
\rho=\frac{I}{d}+\sum_{i=1}^{d^2-1}\alpha_i\frac{A_i}{{\cal{N}}_i},
\end{equation}
where ${\cal{N}}_i={\text{Tr}}\{A_i^2\}$ is the corresponding trace norm and $\alpha_i=\text{Tr}\{\rho A_i\}$.

Let me introduce an operator basis for the qutrit, which is natural for a decomposition into distinct qubit subspaces. It consists of six traceless Gell-Mann matrices \cite{BK} and three diagonal rank $1$ matrices of trace $1$. The first triple of Gell-Mann matrices is symmetric:
\begin{eqnarray}
A_{12} &=& |1\rangle\langle 2|+|2\rangle\langle 1|, \nonumber \\
A_{13} &=& |1\rangle\langle 3|+|3\rangle\langle 1|, \nonumber \\
A_{23} &=& |2\rangle\langle 3|+|3\rangle\langle 2|. \nonumber 
\end{eqnarray}
The second one is antisymmetric  
\begin{eqnarray}
B_{12} &=& -i|1\rangle\langle 2|+i|2\rangle\langle 1|, \nonumber \\
B_{13} &=& -i|1\rangle\langle 3|+i|3\rangle\langle 1|, \nonumber \\
B_{23} &=& -i|2\rangle\langle 3|+i|3\rangle\langle 2|, \nonumber 
\end{eqnarray} 
Both, symmetric and antisymmetric matrices have trace norm equal $2$. The last triple is given by
$$O_1=|1\rangle\langle 1|,~O_2=|2\rangle\langle 2|,~O_3=|3\rangle\langle 3|.$$ All nine operators form an orthogonal operator basis in which one can expand any $3\times 3$ matrix. In particular, the density matrix can be written as
\begin{equation}\label{e2}
\rho=\sum_{i=1}^{3}\left(\omega_i O_i+\sum_{j>i}\left(\frac{\alpha_{ij}}{2} A_{ij} +\frac{\beta_{ij}}{2} B_{ij}\right)\right).
\end{equation}
Throughout the paper it is assumed that for all subscripts $i\neq j \neq k$, and if for some subscript $ij$ $i>j$, then it should be read as $ji$.
The coefficients $\omega_i$, $\alpha_{ij}$ and $\beta_{ij}$ are expectation values of the corresponding operators with respect to $\rho$. It is convenient to write Eq. (\ref{e2}) in an explicit form
$$\rho=\begin{pmatrix} \omega_1 & \frac{\alpha_{12} - i \beta_{12}}{2} & \frac{\alpha_{13} - i \beta_{13}}{2} \\ \frac{\alpha_{12} + i \beta_{12}}{2} & \omega_2 & \frac{\alpha_{23} - i \beta_{23}}{2} \\ \frac{\alpha_{13} + i \beta_{13}}{2} & \frac{\alpha_{23} + i \beta_{23}}{2} & \omega_3 \end{pmatrix}.$$

The three $2\times 2$ principal sub-matrices of the above matrix represent qubit subspaces. It is important to notice that for a positive semidefinite matrix all of its principal sub-matrices are also positive semidefinite (see \cite{HJ}). Thus, principal $2\times 2$ sub-matrices of the qutrit can indeed describe qubits, though unnormalized. The idea is to represent the qutrit by the three Bloch vectors of these qubits. As already mentioned, they cannot be independent, since qutrits are described by only $8$ parameters and each qubit subspace is described by $3+1$ parameters. An additional parameter is due to the lack of the normalization constraint, and the radius of the corresponding Bloch sphere can be less than $1$. For the consistent picture one has to derive the full set of constraints on parameters $\omega_{i}$, $\alpha_{ij}$ and $\beta_{ij}$. 

\section{Constraints on basis expansion coefficients}

First of all, the trace of $\rho$ has to equal $1$, which gives the first constraint
\begin{equation}\label{e3}
\sum_{i=1}^{3}\omega_k=1.
\end{equation}
Since coefficients $\omega_i$ are the smallest principal sub-matrices of $\rho$, they have to be nonnegative. Moreover, due to relation (\ref{e3}) each of them has to obey
\begin{equation}\label{e4}
0\leq \omega_k \leq 1.
\end{equation}
There is nothing new about the above relation, because diagonal elements of density matrix correspond to probabilities of measurement outcomes in the computational basis.

Next, let me find what conditions would provide the positive semi-definiteness of $\rho$. This can be done by studying principal minors with the help of the interlacing inequality \cite{HJ}. This inequality states that if $A$ is an $n\times n$ hermitian matrix and $A'$ is its $(n-1)\times(n-1)$ principal sub-matrix, then the eigenvalues of $A'$ fall exactly in between the eigenvalues of $A$
\begin{equation}\label{e5}
\lambda_1\leq \lambda '_1\leq \lambda_2 \leq \dots \leq \lambda '_{n-1}\leq \lambda_n.
\end{equation}
The eigenvalues are indexed according to a nondecreasing order, i.e., if $i<j$ then $\lambda_i \leq \lambda_j$ (the same holds for $\lambda '_i$). There exists a very nice criterion which states that $A$ is positive definite if and only if all of its principal minors are positive (see \cite{HJ}). In fact, it is enough to consider only leading (upper left) principal minors. Unfortunately, a similar criterion does not work for positive semidefinite matrices. However, below I show that for $3\times 3$ matrices of trace $1$ the matrix $\rho$ is indeed positive semidefinite if and only if all of its principal minors are nonnegative.

If $\rho$ is positive semidefinite, then by the interlacing inequality (\ref{e5}) all of its principal minors are nonnegative. To prove the converse, let me start with $1\times 1$ principal sub-matrices, which are simply the coefficients $\omega_i$. Their non-negativity is already provided by (\ref{e4}). If some of them were negative, then by (\ref{e5}) $\rho$ would have at least one negative eigenvalue. 

If a $2\times 2$ principal minor is negative, then the corresponding $2\times 2$ principal sub-matrix has one positive and one negative eigenvalue. This implies the negativity of at least one eigenvalue of $\rho$, thus no $2\times 2$ principal minor can be negative. If the same minor is positive, the interlacing inequality states that both eigenvalues of the corresponding sub-matrix have to be positive. Finally, if it is zero, one has to consider two cases. If both principal sub-matrices $\omega_i$ and $\omega_j$ are positive, then one eigenvalue is positive and one equals zero. On the other hand, if at least one $\omega_i$ equals zero, then $\alpha_{ij}^2+\beta_{ij}^2$ has to be zero too, but in this case $\alpha_{ij}=\beta_{ij}=0$, thus $\omega_i$ and  $\omega_j$ are simply eigenvalues, which by definition are nonnegative.

The fact that $2\times 2$ principal minors have to be nonnegative can be formulated as three new constraints
\begin{equation}\label{e6}
\alpha_{ij}^2 + \beta_{ij}^2 \leq 4 \omega_i \omega_j, 
\end{equation}
which resemble the well known constraint on the coefficients of the qubit. Interestingly, the above constraints indicates the spherical symmetry on parameters $\alpha_{ij}$ and $\beta_{ij}$.

Since we already know that all principal sub-matrices of $\rho$, whose dimension is smaller than three, have to be positive semidefinite, what is left to study the positive semi-definiteness of $\rho$ is to consider the largest minor, i.e., the determinant of $\rho$. Let me first examine how its sign depends on the eigenvalues of $\rho$. If conditions (\ref{e3}), (\ref{e4}) and (\ref{e6}) are satisfied, then by the interlacing inequality, $\rho$ cannot have more than one negative eigenvalue. Thus, if $\det(\rho)$ is positive, then $\rho$ is positive too. On the other hand, the negativity of $\det(\rho)$ implies that exactly one eigenvalue of $\rho$ is negative. Finally, if $\det(\rho)=0$ the interlacing inequality and condition (\ref{e3}) states that either
\begin{equation}\label{e11}
\lambda_1 \leq 0,~~\lambda_2=0,~~\lambda_3\geq 1,
\end{equation}
or 
\begin{equation}\label{e12}
\lambda_1=0,~~0\leq \lambda_2\leq \lambda_3 \leq 1.
\end{equation}
However, in the first case the trace of the square of $\rho$ should be greater or equal to the trace of $\rho$
$$\text{Tr}(\rho^2)\geq\text{Tr}(\rho).$$
One can easily check that 
\begin{eqnarray}\label{e13}
\text{Tr}(\rho^2)=\sum_{i=1}^3\left(\omega_i^2+\frac{1}{2}\sum_{j>i}
\left(\alpha_{ij}^2+\beta_{ij}^2\right)\right)\leq \nonumber \\
\sum_{i=1}^3\left(\omega_i^2+2\sum_{j>i}
\left(\omega_i\omega_j\right)\right)=\left(\sum_{i=1}^3 \omega_i\right)^2=1,
\end{eqnarray}
therefore only the second case (\ref{e12}) can hold. Thus $\rho$ is positive semidefinite if and only if all of its principal minors are nonnegative. This ends the proof.

To examine the structure of $\det(\rho)$, and additional constraints on $\omega_{i}$, $\alpha_{ij}$ and $\beta_{ij}$, let me consider the three two-dimensional vectors $\vec{v_{ij}}=(\alpha_{ij},\beta_{ij})$, which can be simply interpreted as projections of Bloch vectors onto XY plane. For this purpose it is convenient to introduce the relative length of $\vec{v_{ij}}$
\begin{equation}\label{e7}
d_{ij}=\frac{\sqrt{\alpha_{ij}^2 + \beta_{ij}^2}}{2\sqrt{\omega_i \omega_j}},
\end{equation}
and the angles between $\vec{v_{ij}}$ and axes $A_{ij}$
\begin{equation}\label{e8}
\varphi_{ij}=\arctan\left(\frac{\beta_{ij}}{\alpha_{ij}}\right). 
\end{equation}
The relative length is valid only for nonzero $\omega_i$ and $\omega_j$ and is bounded by $0\leq d_{ij} \leq 1$. Using these new parameters, the determinant of $\rho$ is given by
\begin{equation}\label{e9}
\det(\rho)=\omega_1\omega_2\omega_3\left(1-\sum_{i,j>i}d_{ij}^2+
2d_{12}d_{13}d_{23}\cos\Phi\right),
\end{equation}
where 
\begin{equation}\label{e10}
\Phi=\varphi_{12}-\varphi_{13}+\varphi_{23}.
\end{equation}
Since $\omega_i$ are always nonnegative, one can only consider the inequality
\begin{equation}\label{e14}
1-\sum_{i,j>i}d_{ij}^2+2d_{12}d_{13}d_{23}\cos\Phi \geq 0.
\end{equation}

The relative lengths $d_{ij}$ are related to the pureness of the state. For example, similarly to the qubit case, the state is completely mixed if all $d_{ij}=0$ and all $\omega_i$ are equal. On the other hand, if only one parameter $d_{ij}=1$, the corresponding $2\times 2$ principal minor equals zero, implying $\det(\rho)=0$. The other two principal minors are positive, therefore $\rho$ has to have two positive eigenvalues. In this case the qutrit is confined in a two-dimensional subspace, thus cannot be maximally mixed. Moreover, the left hand side of (\ref{e14}) is less or equal zero, thus the inequality is satisfied only if $\Phi$ is a multiple of $2\pi$ and the other two relative lengths are equal. Finally, if two coefficients $d_{ij}=1$, the third one has to equal $1$ too, and again, $\Phi$ has to be a multiple of $2\pi$. In this case all $2\times 2$ principal minors are $0$. Moreover, $\text{Tr}(\rho^2)=\text{Tr}(\rho)$, which implies that the qutrit is in a pure state. Note, that the pure state of the qutrit is defined by only four independent parameters, which in this case are two $\varphi_{ij}$ and two $\omega_i$. 

The pureness of the qutrit is also dependent on $\Phi$. The parameter $\Phi$ can differ from $k2\pi$ only if all $d_{ij}$ are less than one. Manipulation with relative phases $\varphi_{ij}$ can change the rank of $\rho$ from three to two. From the point of view of vectors $\vec{v_{ij}}$ the dependence on $\Phi$ tells about allowed angles between them. The greater is the relative length, the more dependent on $\Phi$ are the vectors. Some values of $\Phi$ are not allowed for specific $d_{ij}$ due to $(\ref{e14})$. This is an important constraint, because it states that for some relative lengths the vectors cannot point in arbitrary directions. If at least one vector has relative length equal to one, then 
\begin{equation}\label{e15}
\varphi_{12}-\varphi_{13}+\varphi_{23}=0.
\end{equation} 
The set of states for which all values of $\Phi$ are valid is given by the inequality
\begin{equation}\label{e16}
1-\sum_{i,j>i}d_{ij}^2-2d_{12}d_{13}d_{23}\geq 0,
\end{equation}
which is obtained by taking $\cos\Phi=-1$. It is also clear that the dependence on $\Phi$ vanish when at least one relative length is zero.

\section{Three Bloch vectors}

The full set of constraints on $\alpha_{ij}$, $\beta_{ij}$ and $\omega_{i}$ is given by Eqs. (\ref{e3}), (\ref{e4}), (\ref{e6}) and (\ref{e14}). One is able to represent the state of qutrit via three real two-dimensional vectors $\vec{v_{ij}}=(\alpha_{ij},\beta_{ij})$, only if one also gives the values of the maximal allowed ranges $d_{ij}$, or simply the parameters $\omega_i$. However, a three-dimensional Bloch vector description can be used. Let me introduce the three new operators
\begin{equation}\label{e17}
C_{ij}=\frac{-i}{2}[A_{ij},B_{ij}].
\end{equation}
These operators are not linearly independent, since
\begin{equation}\label{e18}
C_{12}-C_{13}+C_{23}=0.
\end{equation}
The expectation value of $C_{ij}$ is equal to
\begin{equation}\label{e19}
\langle C_{ij}\rangle=\omega_i - \omega_j=\gamma_{ij}.
\end{equation}
Due to the commutation relation (\ref{e17}) the three three-dimensional vectors $\vec{u_{ij}}=(\alpha_{ij},\beta_{ij},\gamma_{ij})$ are the Bloch vectors of qubit subspaces, with vectors $\vec{v_{ij}}$ being their projections onto $A_{ij}B_{ij}$ planes. 

The length of $\vec{u_{ij}}$ cannot exceed
\begin{equation}\label{e20}
\sqrt{\alpha_{ij}^2+\beta_{ij}^2+\gamma_{ij}^2} \leq 1-\omega_k = R_{ij},
\end{equation}
where $R_{ij}$ is the radius of the $ij$-th Bloch sphere. This is an additional important constraint. Also note, 
\begin{equation}
\gamma_{ij}=R_{ik}-R_{jk}.
\end{equation}
\begin{figure}
\scalebox{1.0} {\includegraphics[width=8truecm]{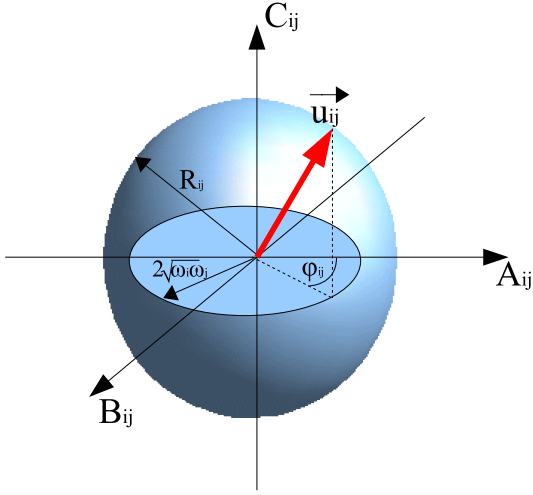}}
\caption{\label{f1} The Bloch vector for a qubit subspace of the qutrit.}
\end{figure}
One of such three vectors is depicted in Fig. \ref{f1}. 

The relations (\ref{e18}-\ref{e20}) lead to yet another two interesting constraints on vectors $\vec{u_{ij}}$
\begin{equation}\label{e21}
\gamma_{12}-\gamma_{13}+\gamma_{23}=0,
\end{equation}
and
\begin{equation}\label{e22}
|\vec{u_{12}}|+|\vec{u_{13}}|+|\vec{u_{23}}| \leq 2.
\end{equation}
The three-dimensional vectors $\vec{u_{ij}}$ are even more dependent
than the two-dimensional ones (namely $\vec{v_{ij}}$), but this is not surprising, since the number of parameters describing the state is constant.
\begin{figure}
\scalebox{1.0} {\includegraphics[width=8truecm]{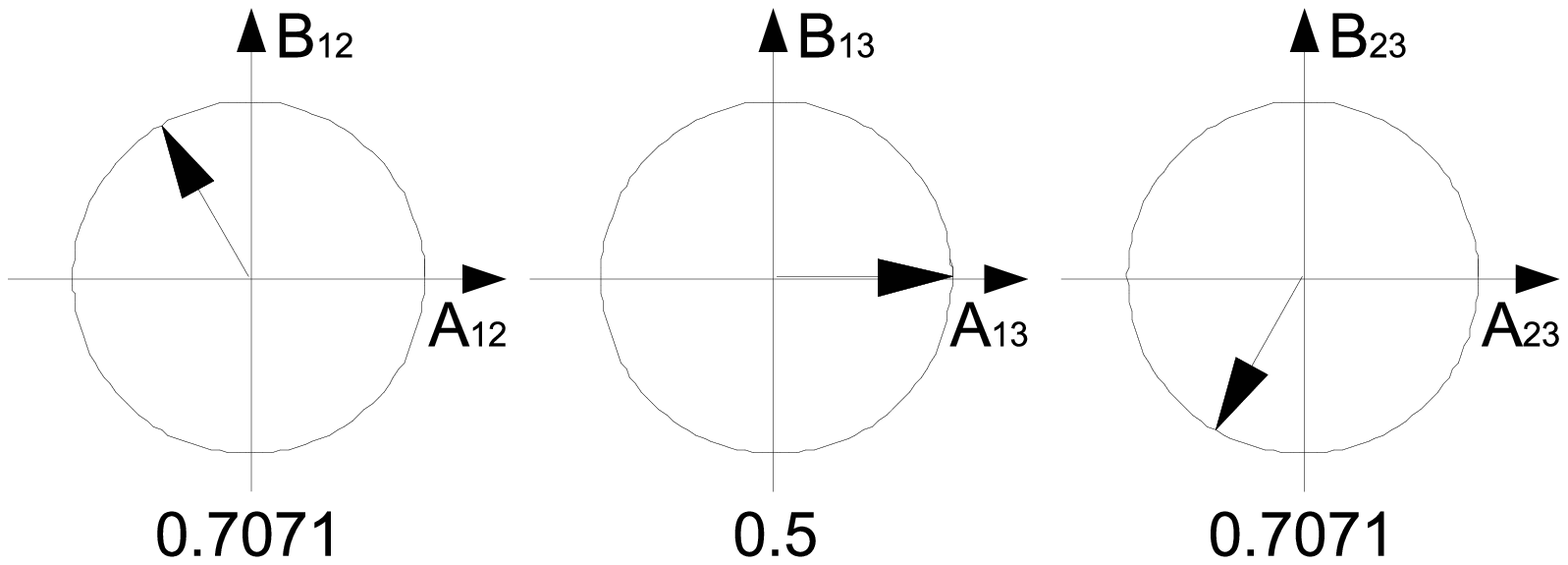}}
\scalebox{1.0} {\includegraphics[width=8truecm]{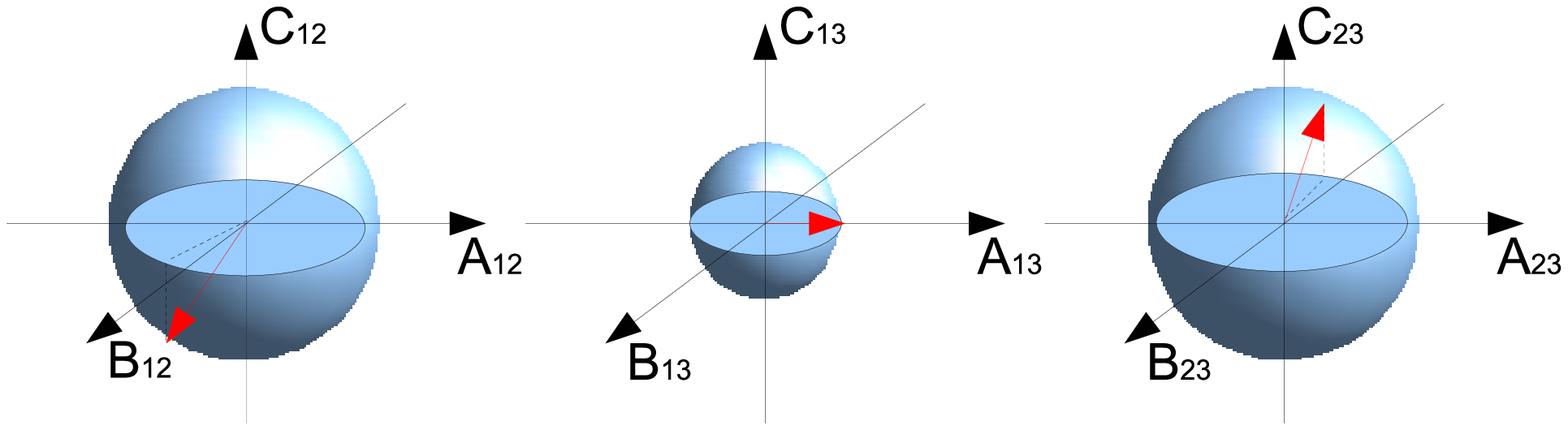}}
\caption{\label{f2} Two representations of the state $(\frac{1}{2},\frac{e^{2\pi i/3}}{\sqrt{2}},\frac{1}{2})^T$. Top --- the two-dimensional representation with normalized vectors, whose length is given in terms of $d_{ij}$, and with explicitly written  maximal lengths $2\sqrt{\omega_i \omega_j}$. Bottom --- a schematic picture of the three Bloch vectors $\vec{u_{ij}}$}
\end{figure}
In fig. \ref{f2} two representations of the state $(\frac{1}{2},\frac{e^{2\pi i/3}}{\sqrt{2}},\frac{1}{2})^T$ are given. The vectors of the two-dimensional representation are normalized, i.e., their length is given in terms of $d_{ij}$.

It is convenient to group all of the constraints on the Bloch vectors and to list them in a compact form:
\begin{equation}
\begin{array}{lll} \sum_{i,j>i} R_{ij} =2,~~0\leq R_{ij} \leq 1, & ~~~~ & (\text{i})  \vspace{1mm} \\ |\vec{u_{ij}}|\leq R_{ij}, & ~~~~ & (\text{ii})  \vspace{1mm} \\
\gamma_{ij}=R_{ik}-R_{jk} & ~~~~ & (\text{iii})  \vspace{1mm} \\
R_{ij}=\frac{1}{3}[2+(-1)^{i+1}\gamma_{ik}+(-1)^j \gamma_{jk}] & ~~~~ & (\text{iv})  \vspace{1mm} \\
\gamma_{12}-\gamma_{13}+\gamma_{23}=0, & ~~~~ & (\text{v})  \vspace{1mm} \\
1-\sum_{i,j>i}d_{ij}^2+2d_{12}d_{13}d_{23}\cos\Phi \geq 0. & ~~~~ & (\text{vi})
\end{array} \nonumber 
\end{equation} 
All the previous constraints on $\alpha_{ij}$, $\beta_{ij}$ and $\omega_{i}$ are contained in the six listed above. This set is somehow over-complete, since some constraints can be derived from the other ones, however I believe that for the clarity it is convenient not to exclude them from the list. 

It is obvious that the set of all Bloch vectors for the qutrit is much richer than the one for the qubit. That is why in the discussion below I will narrow down to examine only the basic properties of the representation. The detailed study of mixed states is not presented here.

\subsection{Pure states and mixed states}

A nice and important property of both 3D and 2D representations is that for pure states all three vectors lie on the corresponding Bloch spheres (circles). This resembles the well known property of the qubit. Moreover, since a pure state is uniquely defined by only four parameters, it is sufficient to describe it by only two vectors, say $\vec{u_{12}}$ and $\vec{u_{13}}$ (alternatively $\vec{v_{12}}$ and $\vec{v_{13}}$ together with $\omega_1$ and $\omega_2$). The third vector is fully dependent on the first and the second one due to $(\text{i-vi})$. 

On the other hand, the completely mixed qubit state is represented by a null vector, i.e. a center of the Bloch sphere. Similarly, in the case of the qutrit one can easily check that in the 3D representation all three vectors can vanish only if all three spheres are of equal radius $2/3$. However, this corresponds exactly to the completely mixed state. As already mentioned, in the 2D representation a degree of mixture is related to the relative length $d_{ij}$. By analogy, in the 3D picture it is related to a ratio $|\vec{u_{ij}}|/R_{ij}$. However, this ratio is not a proper measure of mixedness since it does not include the dependence on the angles $\varphi_{ij}$. 

Another interesting resemblance to the qubit Bloch sphere picture is the fact that if all three vectors are lying along $C_{ij}$ axes, the corresponding states are classical, in the sense that their density matrices are diagonal. Such states correspond to classical mixtures if the computational basis consist of classically allowed states.

\subsection{Orthogonality}

Two states of the qubit are orthogonal if their corresponding Bloch vectors are anti-parallel. For a given pure state of the qubit there is only one uniquely determined orthogonal state. It is not the case for the qutrit, since for an arbitrary pure state there is an orthogonal two-dimensional subspace containing infinitely many states, which are not necessarily pure. Of course, two qutrit states can be orthogonal if at least one of them is pure. One can find an orthogonality criteria by a direct evaluation of $\text{Tr}\{\rho\rho '\}=0$, assuming one density matrix is rank 1, which leads to
\begin{equation}\label{e23}
0=2\sum_{i}\omega_i\omega_i '+ \sum_{i,j>i}\vec{v_{ij}}\cdot\vec{v_{ij}}',
\end{equation}
or equivalently 
\begin{equation}\label{e24}
0=2-\sum_{i,j>i}R_{ij} R_{ij}'+ \sum_{i,j>i}\vec{u_{ij}}\cdot\vec{u_{ij}}'.
\end{equation}
In case both states are pure, the later relation becomes
\begin{equation}\label{e26}
0=2-\sum_{i,j>i}|\vec{u_{ij}}||\vec{u_{ij}}'|\left(1-\cos\theta_{ij}\right),
\end{equation}
where $\theta_{ij}$ is the angle $\vec{u_{ij}}$ makes with $\vec{u_{ij}}'$. Note that for two pure states confined in a two dimensional subspace the above criteria reduce to the qubit orthogonality criterion. For example, if $\omega_3=\omega_3 '=0$ then $R_{12}=R_{12}'=1$ and the pairs of vectors $\vec{u_{13}}$ and $\vec{u_{13}}'$ together with $\vec{u_{23}}$  and $\vec{u_{23}}'$ are pointing in the same direction along $C_{ij}$ axes, thus are parallel. As a result $\cos\theta_{12}=-1$. 

\subsection{Algorithm for valid Bloch vectors}

To summarize this section I will write down a direct algorithm for a construction of any three vectors representing a valid quantum state:
\begin{enumerate}
\item Choose a radius $0 \leq R_{ij} \leq 1$ of an arbitrary Bloch sphere and choose an arbitrary vector $\vec{u_{ij}}=(\alpha_{ij},\beta_{ij},\gamma_{ij})$ lying inside this sphere. (As a consequence, all remaining $R_{ij}$ and $\gamma_{ij}$ are fixed --- constraints $(\text{i-iv})$)
\item Choose the lengths of the remaining two vectors, so that they are confined in the corresponding Bloch spheres.
\item Choose angles between the projections $\vec{v_{ij}}$ and the corresponding $A_{ij}$ axes, so that $(\text{vi})$ is satisfied.
\end{enumerate}

\section{Dynamics}

In the following section I am going to consider the unitary evolution of the qutrit from the point of view of the three Bloch vectors $\vec{u_{ij}}$.
The dynamics of the qutrit is given by eight $SU(3)$ generators, which in our case can be taken to be $A_{ij}$, $B_{ij}$ and two of the three $C_{ij}$ operators. Any unitary operation on the qutrit can be decomposed into a product of unitary operations $U_G(\theta)=e^{iG\theta}$, where $G$ is one of eight generators and $\theta$ is a real parameter. Therefore, it is enough to examine actions of these operators on the three Bloch vectors. To do this, it is convenient to introduce a table of commutation rules for the above operators (see Table 1). 
\begin{table}[p] 
\begin{tabular}{c||c|c|c|c|c|c|c|c|c}
  & $A_{12}$ & $B_{12}$ & $C_{12}$ & $A_{13}$ & $B_{13}$ & $C_{13}$ & $A_{23}$ & $B_{23}$ & $C_{23}$ \\
\hline
\hline
$A_{12}$ & $0$ & $2iC_{12}$ & $-2iB_{12}$ & $iB_{23}$ & $-iA_{23}$ & $-iB_{12}$ & $iB_{13}$ & $-iA_{13}$ & $iB_{12}$\\
\hline
$B_{12}$ & $-2iC_{12}$ & $0$ & $2iA_{12}$ & $iA_{23}$ & $iB_{23}$ & $iA_{12}$ & $-iA_{13}$ & $-iB_{13}$ & $-iA_{12}$ \\
\hline
$C_{12}$ & $2iB_{12}$ & $-2iA_{12}$ & $0$ & $iB_{13}$ & $-iA_{13}$ & $0$ & $-iB_{23}$ & $iA_{23}$ & $0$\\
\hline
$A_{13}$ & $-iB_{23}$ & $-iA_{23}$ & $-iB_{13}$ & $0$ & $2iC_{13}$ & $-2iB_{13}$ & $iB_{12}$ & $iA_{12}$ & $-iB_{13}$ \\
\hline
$B_{13}$ & $iA_{23}$ & $-iB_{23}$ & $iA_{13}$ & $-2iC_{13}$ & $0$ & $2iA_{13}$ & $-iA_{12}$ & $iB_{12}$ & $iA_{13}$ \\
\hline
$C_{13}$ & $iB_{12}$ & $-iA_{12}$ & $0$ & $2iB_{13}$ & $-2iA_{13}$ & $0$ & $iB_{23}$ & $-iA_{23}$ & $0$\\
\hline
$A_{23}$ & $-iB_{13}$ & $iA_{13}$ & $iB_{23}$ & $-iB_{12}$ & $iA_{12}$ & $-iB_{23}$ & $0$ & $2iC_{23}$ & $-2iB_{23}$ \\
\hline
$B_{23}$ & $iA_{13}$ & $iB_{13}$ & $-iA_{23}$ & $-iA_{12}$ & $-iB_{12}$ & $iA_{23}$ & $-2iC_{23}$ & $0$ & $2iA_{23}$ \\
\hline
$C_{23}$ & $-iB_{12}$ & $iA_{12}$ & $0$ & $iB_{13}$ & $-iA_{13}$ & $0$ & $2iB_{23}$ & $-2iA_{23}$ & $0$ \\
\end{tabular}
\caption{Commutation rules of $A_{ij}$, $B_{ij}$ and $C_{ij}$.}
\end{table} 

Let me start with the operators $C_{ij}$, which generate rotations of the vectors $\vec{u_{ij}}$ about $C_{ij}$ axes by the angle $2\theta$. In addition, they also rotate vector $\vec{u_{ik}}$ about $C_{ik}$ and $\vec{u_{jk}}$ about $C_{jk}$ by the angle $\theta$. However, for $C_{12}$ and $C_{23}$, the rotation of $\vec{u_{23}}$ and $\vec{u_{12}}$, respectively, is done in the opposite direction (by $-\theta$).

In addition to the three standard cyclic commuting triples $\{A_{ij},B_{ij},C_{ij}\}$, for which the commutator of the two subsequent operators gives the third operator multiplied by $2i$, there are four extra cyclic commuting triples: $\{A_{12},A_{13},B_{23}\}$, $\{A_{12},A_{23},B_{13}\}$, $\{A_{13},A_{23},B_{12}\}$ and $\{B_{12},B_{13},B_{23}\}$, for which the commutator of the two subsequent operators gives the third operator multiplied only by $i$. Due to the cyclic commutation relation, the mean values of the operators in each triple can be considered as coordinates of yet another Bloch-like vector, for which the operators $A_{ij}$ and $B_{ij}$ are generators of simple rotations. 

From the point of view of the already introduced Bloch vectors $\vec{u_{ij}}$, the generators $A_{ij}$ and $B_{ij}$ cause an oscillation of the coordinates of $\vec{u_{ik}}$ and $\vec{u_{jk}}$. However, a unitary operation cannot change a pure state into a mixed state, and vice versa, therefore, oscillations of coordinates of two vectors have to occur together with simultaneous oscillations of the Bloch sphere radii $R_{ik}$ and $R_{jk}$. This will make pure states having their Bloch vectors always on the surface of the corresponding Bloch spheres. Indeed, $A_{ij}$ generates the following change of the Bloch sphere radii
\begin{eqnarray}
R_{ik}(\theta) & = & \frac{1}{2}\left(1+\omega_k + {\cal A}_{ij}\sin(2\theta + \phi)\right), \label{e27} \\
R_{jk}(\theta) & = & \frac{1}{2}\left(1+\omega_k - {\cal A}_{ij}\sin(2\theta + \phi)\right). \label{e28}
\end{eqnarray}
The action of $B_{ij}$ is similar
\begin{eqnarray}
R_{ik}(\theta) & = & \frac{1}{2}\left(1+\omega_k + {\cal B}_{ij}\cos(2\theta + \phi)\right), \label{e29} \\
R_{jk}(\theta) & = & \frac{1}{2}\left(1+\omega_k - {\cal B}_{ij}\cos(2\theta + \phi)\right). \label{e30}
\end{eqnarray}
In the above, ${\cal A}_{ij}=\sqrt{\beta_{ij}^2+\gamma_{ij}^2}$, ${\cal B}_{ij}=\sqrt{\alpha_{ij}^2+\gamma_{ij}^2}$ and $\phi$ is a phase factor.

The unitary evolution of the qutrit is given by a rotation of the Bloch vectors $\vec{u_{ij}}$ and a resizing of the Bloch spheres. It is more complex than the simple evolution of the qubit, however in this case the number of generators is almost three times larger than in the qubit case. In the next section I will show that these generators have a natural interpretation for spin 1.

\section{Spin 1}

Spin 1 matrices $S_x$, $S_y$, and $S_z$ are cyclic commuting operators $[S_i,S_j]=i\varepsilon_{ijk}S_k$ whose eigenvalues are $-1$, $0$ and $1$. The first resemblance to the operators $A_{ij}$ and $B_{ij}$ is that both, spin 1 matrices and $A_{ij}$ and $B_{ij}$, have the same spectra. Moreover, the same cyclic commutation relations of some triples of $A_{ij}$ and $B_{ij}$ matrices were presented in the previous section. Indeed, if one represents spin 1 matrices in a basis spanned by their eigenvectors corresponding to the eigenvalue zero (the three vectors $|S_{i}=0\rangle$, $i=x,y,z$, are orthogonal), one obtains $A_{23}$, $B_{12}$ and $A_{13}$ respectively. In this case $O_i=|S_{i}=0\rangle\langle S_{i}=0|$.

Next, note that squares of the operators of the $ij$-th Bloch sphere obey the following relation
\begin{equation}\label{e31}
A_{ij}^2=B_{ij}^2=C_{ij}^2=I-O_k.
\end{equation}
As a consequence, $\langle B_{12}^2 \rangle+\langle A_{13}^2 \rangle+\langle A_{23}^2 \rangle=R_{12}+R_{13}+R_{23}=2$. On the other hand, the well known angular momentum formula yields that $\langle S_x^2 \rangle+\langle S_y^2 \rangle+\langle S_z^2 \rangle=s(s+1)$, where $s$ is the spin number, which for spin 1 gives the right hand side equal to $2$. The above seems to lead to the following interpretation of the three Bloch spheres: each Bloch sphere describes a different spatial coordinate of spin 1 and the sphere radius corresponds to $\langle S_i^2 \rangle$. The question is, what is the meaning of the other two Bloch sphere coordinates orthogonal to the one denoting $S_i$?

The operators $C_{ij}=O_i - O_j$, therefore $C_{ij}=|S_{i}=0\rangle\langle S_{i}=0|-|S_{j}=0\rangle\langle S_{j}=0|=S_j^2-S_i^2$. A difference of squares of two spin matrices is a generator of a certain type of spin squeezing known as two-axes countertwisting \cite{KU}. In fact, the remaining three operators $A_{12}$, $B_{13}$ and $B_{23}$ are also generators of two-axes countertwisting. Note, that they can be obtained from the first triple as anti-commutatators $\{B_{12},A_{13}\}=B_{12}A_{13}+A_{13}B_{12}=-B_{23}$, $\{B_{12},A_{23}\}=B_{13}$ and $\{A_{13},A_{23}\}=A_{12}$. To see that anti-commutators of spin matrices are two-axes countertwisting generators, let me write them explicitly
\begin{eqnarray}
\{S_i,S_j\} & = & S_i S_j + S_j S_i = \frac{S_i + S_j}{\sqrt{2}}\frac{S_i + S_j}{\sqrt{2}} \nonumber \\ 
 &-& \frac{S_i - S_j}{\sqrt{2}}\frac{S_i - S_j}{\sqrt{2}}= S_{i+j}^2-S_{i-j}^2,\nonumber
\end{eqnarray}
where $S_{i+j}$ and $S_{i-j}$ are spin matrices for the orthogonal directions $i+j$ and $i-j$.

The physical meaning of coordinates of a single Bloch vector for spin 1 are mean  values of the following observables: spin coordinate $S_i$, generator of two-axes countertwisting along two orthogonal axes $j$ and $k$ (which are both orthogonal to $i$), and generator of two-axes countertwisting along axes $j+k$ and $j-k$. At this point, let me discuss what information one gains while measuring the last two observables. The eigenvectors of a two-axes countertwisting generator $S_{i}^2-S_{j}^2$ are $|S_{i}=0\rangle$, $|S_{j}=0\rangle$ and $|S_{k}=0\rangle$, where directions $i$, $j$ and $k$ are mutually orthogonal. If the measurement yields an outcome $|S_{i}=0\rangle$, one knows that the mean spin vector is definitely {\it not} lying along $i$ axis, therefore the measurement of a two-axes countertwisting generator corresponds to the question: along which one of the three mutually orthogonal directions $i$, $j$ and $k$ the spin vector is definitely not lying (see \cite{KKC}). 

Finally, one also sees that every unitary transformation of spin 1 can be represented as a combination of three rotations and six two-axes countertwistings (however, it is enough to consider only five of them, since $C_ij$ are linearly dependent). Moreover, an intuitive interpretation of rotations and squeezing makes the change of the Bloch sphere radii quite natural, since rotations interchange spatial coordinates and squeezing deforms them.

Similarly to the qubit case, the proposed Bloch vector representation of the qutrit has natural interpretation within the framework of the spin. It is an open question whether one can obtain similar property for the Bloch vector representation of higher dimensional systems.

\section{Higher dimensions}

An analogical operator basis to the one used in this work can be easily constructed for any d-dimensional system. In such a basis one can easily represent any state by a collection of Bloch vectors. One may also try to find an analogy between basis operators and spin operators, however the basis operators are either rank one projectors or Pauli matrices acting in a two dimensional subspace, thus have eigenvalues $1$, $-1$ and $0$, which is $d-2$ times degenerated. On the other hand, spin operators, whose eigenvalues are $-\frac{2d-1}{2},-\frac{2d-3}{2},\dots,\frac{2d-1}{2}$, are not degenerated and act on the whole space. Yet, they can be written as a direct sum of the operators proportional to the Pauli matrices. In case $d$ is odd, one also has to include $0$ in the direct sum expansion, just like it is done for the qutrit.

It is much harder to show what constraints are to be satisfied by the collection of Bloch vectors to properly represent quantum states. It is obvious that the interlacing inequality requires all principal sub-matrices of the density matrix to be positive semidefinite. This implies that all principal minors of $\rho$ are nonnegative. Unfortunately, for general density matrices the converse may not hold. Nevertheless, the non-negativity of principal minors gives necessary conditions for the positive semi-definiteness of $\rho$. Therefore, there are at least $\sum_{n=0}^{d} \begin{pmatrix} d \\ n \end{pmatrix}=2^d$ constraints on the basis expansion coefficients, where the case $n=0$ corresponds to the normalization constraint. The exponential number of constraints is in stark contrast to the quadratic number of qubit subspaces. Due to this fact, the higher is the level of the system, the more dependent the corresponding Bloch vectors are. There are $d^2-1$ parameters characterizing the state of the system and the average number of parameters per Bloch vector is $\frac{2(d+1)}{d}$.

The multi-vector representation can be especially useful for pure states, which are described by $2(d-1)$ parameters. In such case, the average number of parameters per Bloch vector is $\frac{4}{d}$ and one may expect the state to be representable by fever vectors. In particular, if it was possible to encode two parameters per vector, $d-1$ vectors would be enough to uniquely represent the state.

\section{Conclusions}

I have introduced a Bloch-like representation of the qutrit, which uses three vectors instead of one. The full set of constraints on vectors has been derived (Eqs. i-vi) and it was shown that many properties of the representation resemble the ones for the qubit. In particular, pure states correspond to vectors lying on the surface of Bloch spheres and mixed states to vectors lying inside. Moreover, the dynamics of the qutrit can be described by rotations of the Bloch vectors and a change of the Bloch sphere radii. The generators of these transformations have natural interpretation for spin 1, namely, they generate rotations and two-axes countertwisting spin squeezing. I also discuss basic properties of a generalization of this representation to higher dimensions.

I acknowledge enriching discussions with Antoni W{\'o}jcik. This work was supported from Tomasz {\L}uczak's subsidium MISTRZ (Foundation for Polish Science) and the Scientific Scholarship of the City of Pozna{\'n}.

\end{document}